\documentclass[11pt,twoside]{article} 
\usepackage{graphicx}
\usepackage{amsmath}

\begin{document} 
\thispagestyle{empty}

\markboth{ ~~\hrulefill~~  Bayesian estimation of discrete Burr distribution with two parameters }{}

\topmargin=0mm
\vspace{3.5cm}
{\Large \bf \noindent Bayesian estimation of discrete Burr distribution with two parameters}\\[1cm]

{\centering {\bf Halaleh Kamari and Hossein Bevrani and Kaniav Kamary}\\  
Paris-Sud University, France\\ 
University of Tabriz, Iran\\
Universit{\'e} Paris-Dauphine\\}

\vspace{1cm}
{\noindent \bf Abstract.} 
So far, various techniques have been implemented for generating discrete distributions based on continuous distributions. The characteristics and properties of this kind of probability distributions have been studied. Furthermore, the estimation of related parameters have been computed trough classical methods. However, a few studies addressed the parameter estimate issue of these distributions through Bayesian methods. This is essentially because of the complexity of the model whatever the number of parameter is and the fact that in general they contain a large number of parameters to be estimated. This paper deals with computing Bayes estimate of the parameters of discrete Burr distribution with two parameters. 
Since the resulting posterior distribution of the parameters is not standard, we apply Metropolis-Hastings algorithm to simulate from the posterior density.
 \vspace*{1cm}


\section{Introduction}
In statistics and probability theory, a discrete probability distribution is a distribution characterized by a probability mass function. The discrete random variables have attracted the attention of many researchers due to their applications in many fields. One of them is their great importance to model real-life scenarios. Also, much of the information bioinformatics deals with is discrete data
 and sequence information is usually analyzed using discrete random variables. 

In reliability lifetime modeling, it is common to treat failure data as being continuous, implying some degree of precision in measurement. Too often in practice, however, failures are either noted at regular inspection intervals, occur in a discrete process or are simply recorded in bins. An important aspect of lifetime analysis is to find a lifetime distribution that can adequately describe the aging behavior of the device concerned.
Howevere, it is sometimes impossible or inconvenient to measure the life length of a device on a continuous scale. Thus, it is essential to construct a discrete lifetime models for discrete failure time models.
Various methods of constructing discrete lifetime models have been developed by discretizations of continuous lifetime models, in particular, through discretizations of their hazard rates. 
Using these techniques, continuous distributions can be transformed to discrete distributions. Katz \cite{11} , Roknabadi et al. \cite{22}, Krishna and Pundir \cite{Krishna} have expressed different methods to generate discrete families of distributions. Burr distribution with two parameters is obtained using the general approach of discretizing continuous distribution which can be considered as suitable lifetime models. In this work, we are interested in the parameter estimation issue for the Burr distribution. From a classical point of view, the parameters are supposed to be fixed and the classical estimation methods such as Maximum Likelihood and Moment Method \cite{Parsian} can be used. On the other hand from the Bayesian perspective, the parameters of the model of interest are random variables. In this case, Bayesian methods are applied in order to estimate the parameters. 
After specifying the prior distribution for the parameters, since the resulting posterior distribution is  non-standard, Metropolis-Hastings algorithm is applied in order to simulate from the posterior distribution. The results of implementing the Metropolis-Hastings algorithm are summarized by averaging over the draws in the case of supposing a squared loss function and it is considered as the approximations of the parameter estimators. We proceed here with a briefly review of some distributional properties of Burr distribution.
\\
A continuous lifetime random variable $X$ follows the Burr-XII distribution Br($\alpha$, $\beta$) if its probability density function is as follows:
\begin{equation*}
f(x)=\frac{\alpha\beta x^{\alpha-1}}{(1+x^\alpha)^{\beta+1}},\quad x>0,\alpha,\beta>0
\end{equation*}
For every  $x>0$, the survival function $S(x)$ and the failure rate $r(x)$, that is defined as the ratio of the probability density to the failure rate, are respectively given by
\begin{equation*}
S(x)=(1+x^\alpha)^{-\beta}; \quad r(x)=\frac{\alpha\beta x^{\alpha-1}}{1+x^\alpha}
\end{equation*}
The Second rate of failure is also given by
\begin{equation*}
r^{\ast}(x)=log[S(x)/S(x+1)]=-\beta log[\frac{1+x^\alpha}{1+(1+x)^\alpha}]
\end{equation*}
and the $r$'s moment is
\begin{equation*}
E(X^r)=\beta B(r\alpha^{-1}+1,\beta-r\alpha^{-1}),\quad \alpha,\beta>0,\alpha\beta>r>0
\end{equation*}
where $B(a,b)=\int_0^1 x^{a-1} (1-x)^{b-1} dx$. In a particular case where $\alpha=1$, we get Pareto distribution $Par(\beta)$ \footnote{continuous Pareto distribution with parameter $\beta$} with its own reliability characteristics. If times are grouped into unit intervals, the discrete observed variable is $dX=[X]$ which is considered as the largest integer less than or equal to $X$. It will have the probability function as following:
\begin{equation}\label{b2}
P(dX=x)=p(x)=P(x\leq{X}<x+1)=S(x)-S(x+1) 
\end{equation}
where $x=0,1,2,\dots$. Krishna and Pundir \cite{Krishna} applied this method to study the discrete Pareto and discrete Burr distributions. Using equation (\ref{b2}), we can define the probability mass function of discrete Burr distribution, namely $DBD(\alpha, \theta)$ \footnote {discrete Burr distribution with parameters $\alpha$ and $\theta$}, as following:
\begin{equation*}
p(x)=\theta^{log(1+x^\alpha)}-\theta^{log\{1+(1+x)^\alpha\}},\quad x=0,1,2,...
\end{equation*}
Note that the failure rate value $S(x)$ is the same for both $Br(\alpha,\beta)$ and $DBD(\alpha,\theta)$ in the cases where the variable $x$ takes the integer values. The next section focuses on the Bayesian estimation of the parameters of $DBD$, $\alpha$ and $\theta$.\\

\section{Bayesian estimation of parameters}\label{sec:zero}
Suppose that $X_1, ..,X_n$ are from a discrete Burr distribution with parameters $\alpha$ and $\theta$, then
\begin{equation}\label{d26}
p_{\alpha,\theta}(\underline{x})=\Pi_{i=1}^n(\theta^{log(1+{x_i}^\alpha)}-\theta^{log[1+(1+{x_i})^\alpha]}),x_i=0,1,...,0<\theta<1,\alpha>0
\end{equation}
which is equivalent to 
\begin{equation}\label{d29}
p_{\theta}(\underline{x})\propto{\Pi_{i=1}^n(\theta^{w_{i1}}-\theta^{w_{i2}})}
\end{equation}
where $w_{i1}=log(1+{x_i}^\alpha)$ and $w_{i2}=log[1+(1+{x_i})^\alpha]$. We consider two following cases. 
\subsection{$\alpha$ or $\theta$ is unknown}
When $\alpha$ is known, because $\theta$ belongs to the unit interval, a natural prior distribution choice can be a beta distribution with hyperparameters $a$ and 1 that is written as
\begin{equation}\label{d28}
\pi(\theta)\propto{\theta^{a-1}}
\end{equation}
The posterior distribution of $\theta$ is therefore obtained as follows:
\begin{align}
\pi(\theta|\underline{x})&\propto{[\Pi_{i=1}^n(\theta^{w_{i1}}-\theta^{w_{i2}})]\theta^{a-1}}\nonumber\\
&\propto{\theta^{a+\sum_{i=1}^nw_{i1}-1}[\Pi_{i=1}^n(1-\theta^{w_i})]}
\end{align}
By considering $w_i=w_{i2}-w_{i1}$, we will have%
\begin{equation}\label{d32}
\pi(\theta|\underline{x})=\left(\Pi_{i=1}^n(\frac{w_i}{\delta_i\tau_i})\right)^{-1}\theta^{a+\sum_{i=1}^nw_{i1}-1}[\Pi_{i=1}^n(1-\theta^{w_i})]
\end{equation}
By assuming a squared-error loss function the Bayesian estimation of the parameter $\theta$ can easily be obtained as:
\begin{equation}\label{d36}
\theta^{*}=\frac{1}{\Pi_{i=1}^n(\frac{w_i}{\delta_i\tau_i})}\Pi_{i=1}^n(\frac{w_i}{\lambda_i\rho_i})
\end{equation}
where $\lambda_i=w_{i1}+\frac{a}{n}+1$, $\rho_i=w_{i2}+\frac{a}{n}+1$, $\delta_i=w_{i1}+\frac{a-1}{n}+1$ and $\tau_i=w_{i2}+\frac{a-1}{n}+1$.\\
While only $\alpha$ is assumed to be unknown we consider the case where there is no a priori information about the distribution of $\alpha$. We therefore define the following prior density for $\alpha$:
\begin{equation}\label{h7}
\pi(\alpha)\propto{\frac{1}{\alpha}},\alpha>0
\end{equation}
The posterior distribution obtained by combining the likelihood with the prior distribution will be
\begin{equation}
\pi(\alpha|\underline{x})\propto{\frac{1}{\alpha}\theta^{\sum_{i=1}^nw_{i1}}[\Pi_{i=1}^n(1-\theta^{w_i})]}
\end{equation}
which implies that
\begin{equation}\label{h10}
\pi(\alpha|\underline{x})=Z(\theta)\frac{1}{\alpha}\theta^{\sum_{i=1}^nw_{i1}}[\Pi_{i=1}^n(1-\theta^{w_i})]
\end{equation}
where $Z(\theta)^{-1}$ is the unconditional marginal distribution of the random variable $X$ and is calculated as follows:
\begin{equation}\label{d41}
Z^{-1}(\theta)=\int_0^{\infty}\frac{1}{\alpha}\theta^{\sum_{i=1}^nw_{i1}}[\Pi_{i=1}^n(1-\theta^{w_i})]d\alpha
\end{equation}
Since the posterior density obtained in (\ref{h10}) is not standard, we use the numerical methods by programing in R software. A simulation study involves three sample of size $25$ that are simulated from the discrete Burr distribution when the true values of $\theta$ and $\alpha$ are $0.1, 0.2, 0.3$ and $1, 2, 3$, respectively. We implement the Metropolis-Hastings algorithm \cite{Robert} by considering a gamma distribution as the default proposal distribution for the parameter $\alpha$. Tables \ref{pd1} and \ref{pd2} show the summary statistics of $10^4$ MCMC iterations based on squared and absolute error loss functions. By comparing the true values of $\alpha$ noted in the first row of the tables with the resulting estimations, Tables \ref{pd1} and \ref{pd2} illustrate the accuracy of the method to compute the posterior estimates of $\alpha$ in this case. \\
\begin{table}‬
\begin{center}
\caption{\small{
Simulation results when the loss function is squared error.}}\label{pd1}
\begin{scriptsize}
\begin{tabular}{|cc|c|c|c|c|}
\hline
$\theta\downarrow$ &  & $\alpha$=1 & $\alpha$=2 & $\alpha$=3 & $\alpha$=4   \\
\hline
0.1 & $\hat{\alpha}$ & 1.2120 & 1.9623 & 2.8210 & 3.9420\\
 & $var(\hat{\alpha})$ & 0.0829 & 0.2222 & 0.5887 & 0.6661 \\
 \hline
0.2 & $\hat{\alpha}$ & 1.0285 & 2.0540 & 3.0560 & 3.9910\\
 & $var(\hat{\alpha})$ & 0.1609 & 0.4266 & 1.1023 & 1.2491\\
\hline
0.3 & $\hat{\alpha}$ & 1.0747 & 2.0355 & 3.0020 & 4.0100\\
 & $var(\hat{\alpha})$ & 0.2546 & 0.6886 & 1.7974 & 1.9379\\
\hline
\end{tabular} 
\end{scriptsize}
\end{center}
\end{table}‬

\begin{table}‬
\begin{center}
\caption{\small{
Simulation results when the loss function is absolute error.}}\label{pd2}
\begin{scriptsize}
\begin{tabular}{|cc|c|c|c|c|}
\hline
 $\theta\downarrow$ &  & $\alpha$=1 & $\alpha$=2 & $\alpha$=3 & $\alpha$=4  \\
\hline
0.1 & $\hat{\alpha}$ & 1.1900 & 1.9096 & 2.7200 & 2.8180\\
 & $var(\hat{\alpha})$ & 0.0829 & 0.2222 & 0.5887 & 0.6661\\
 \hline
 0.2 & $\hat{\alpha}$ & 1.5926 & 2.5727 & 3.7270 & 3.8140 \\
 & $var(\hat{\alpha})$ & 0.1609 & 0.4266 & 1.1023 & 1.2491 \\
\hline
 0.3 & $\hat{\alpha}$ & 2.0294 & 3.2620 & 4.7120 & 4.7900 \\
 & $var(\hat{\alpha})$ & 0.2546 & 0.6886 & 1.7974 & 1.9379\\
\hline
\end{tabular} 
\end{scriptsize}
\end{center}
\end{table}‬

\subsection{$\alpha$ and $\theta$ are both unknown}
Joint prior density of $\alpha$ and $\theta$ can be as following
\begin{equation}\label{d37}
\pi(\alpha,\theta)\propto{\frac{1}{\alpha}\theta^{a-1}},0<\theta<1,\alpha>0
\end{equation}
which results in the following joint posterior distribution 
\begin{equation}\label{d40}
\pi(\alpha,\theta|\underline{x})=\left(\int_0^{\infty}\frac{1}{\alpha}\Pi_{i=1}^n(\frac{w_i}{\delta_i\tau_i})(\alpha)d\alpha\right)\frac{1}{\alpha}\theta^{a+\sum_{i=1}^nw_{i1}-1}[\Pi_{i=1}^n(1-\theta^{w_i})]
\end{equation}
where $w_i=w_{i2}-w_{i1}$, $\delta_i=w_{i1}+\frac{a-1}{n}+1$ and $\tau_i=w_{i2}+\frac{a-1}{n}+1$.\\
In order to simulate from the posterior distribution above we apply the Metropolis-Hastings algorithm when the candidate proposals of $\alpha$ and $\theta$ are simulated from a uniform U$(0, 1)$ and a gamma distributions centered on the maximum likelihood estimate of $\theta$, respectively. By running the algorithm with $10^4$ MCMC iterations for three datasets described in the previous section, the posterior estimations of the parameters are summarized in Tables \ref{pd3} and \ref{pd4}. Once again, both parameters $\alpha$ and $\theta$ are accurately estimated while the variance of the posterior estimates are at least $0.08$.
\begin{table}‬
\begin{center}
\caption{\small{
Simulation results when the loss function is squared error.}}\label{pd3}
\begin{scriptsize}
\begin{tabular}{|cc|c|c|c|c|}
\hline
$\theta\downarrow$ &  &  $\alpha$=1 & $\alpha$=2 & $\alpha$=3 & $\alpha$=4   \\
\hline
0.1 & $\hat{\alpha}$ & 1.1002 & 1.9120 & 2.920 & 3.644 \\
 & $\hat{\theta}$ & 0.0812 & 0.0965 & 0.1592 & 0.1379\\
 & $var(\hat{\alpha})$ & 0.5784 & 11.7378 & 36.1284 & 55.2465\\
 & $var(\hat{\alpha})$ & 0.5784 & 11.7378 & 36.1284 & 55.2465\\
 & $var(\hat{\theta})$ & 0.0069 & 0.0114 & 0.0059 & 0.0029\\
 & $corr(\hat{\alpha},\hat{\theta})$ & 0.2046 & 0.3094 & 0.3242 & 0.2523 \\
 \hline
0.2 & $\hat{\alpha}$ & 1.1918 & 2.4980 & 2.7800 & 3.9300\\
 & $\hat{\theta}$ & 0.1915  & 0.1742 & 0.2180 & 0.1728\\
 & $var(\hat{\alpha})$ & 0.5880 & 14.1243 & 13.2204 & 32.2304\\
 & $var(\hat{\theta})$ & 0.0102 & 0.0071 & 0.0039 & 0.0019\\
 & $corr(\hat{\alpha},\hat{\theta})$ & 0.0557 & 0.2473 & 0.1675 & 0.1735\\ 
\hline
0.3 & $\hat{\alpha}$ & 1.6080 & 2.1800 & 3.5700 & 3.7600\\
 & $\hat{\theta}$ & 0.2701 & 0.2564 & 0.3121 & 0.3422\\
 & $var(\hat{\alpha})$ & 2.2266 & 18.6511 & 32.4364 & 33.6526\\
 & $var(\hat{\theta})$ & 0.0083 & 0.0028 & 0.0027 & 0.0034\\
 & $corr(\hat{\alpha},\hat{\theta})$ & 0.1066 & 0.1883 & 0.2394 & 0.2615\\
\hline
\end{tabular} 
\end{scriptsize}
\end{center}
\end{table}‬

\begin{table}‬
\begin{center}
\caption{\small{
Simulation results when the loss function is absolute error. }}\label{pd4}
\begin{scriptsize}
\begin{tabular}{|cc|c|c|c|c|}
\hline
 $\theta\downarrow$ &  & $\alpha$=1 & $\alpha$=2 & $\alpha$=3 & $\alpha$=4 \\
\hline
0.1 & $\hat{\alpha}$ & 1.0970 & 1.8170 & 3.1910 & 3.9790\\
 & $\hat{\theta}$ & 0.0817 & 0.1022 & 0.0703 & 0.1451\\
 & $var(\hat{\alpha})$ & 0.5784 & 11.7378 & 36.1284 & 55.2465\\ 
 & $var(\hat{\theta})$ & 0.0069 & 0.0114 & 0.0059 & 0.0029\\
 & $corr(\hat{\alpha},\hat{\theta})$ & 0.2046 & 0.3094 & 0.3242 & 0.2523\\
 \hline
0.2 & $\hat{\alpha}$ & 1.0560 & 1.8940 & 2.6600 & 4.000\\
 & $\hat{\theta}$ & 0.2918 & 0.2563 & 0.1966 & 0.1810\\
 & $var(\hat{\alpha})$ & 0.5880 & 14.1243 & 13.2204 & 32.2304\\
 & $var(\hat{\theta})$ & 0.0102 & 0.0071 & 0.0039 & 0.0019\\
 & $corr(\hat{\alpha},\hat{\theta})$ & 0.0557 & 0.2473 & 0.1675 & 0.1735\\
\hline
0.3  & $\hat{\alpha}$ & 1.2630 & 1.8600 & 3.3100 & 3.6600\\
 & $\hat{\theta}$ & 0.3068 & 0.2681 & 0.2725 & 0.3207\\
 & $var(\hat{\alpha})$ & 2.2266 & 18.6511 & 32.4364 & 33.6526\\
 & $var(\hat{\theta})$ & 0.0083 & 0.0028 & 0.0027 & 0.0034\\
 & $corr(\hat{\alpha},\hat{\theta})$ & 0.1066 & 0.1883 & 0.2394 & 0.2615\\
\hline
\end{tabular} 
\end{scriptsize}
\end{center}
\end{table}‬

\section{Conclusion}\label{sec:quatr}
This paper deals with computing Bayesian estimation of the parameters of discrete Burr distribution with two parameters by applying the Metropolis-Hastings algorithm. Simulated datasets have been used in order to compute the parameter estimates based on two loss functions, squared-error and absolute-error. While the posterior distribution of the parameters is non-standard, the MCMC results illustrate that this method is satisfactory to estimate the unknown parameters of the discrete Burr distribution. \\

\end{document}